\documentstyle[11pt]{article}

\newcommand{\ome}{\bar{\Omega}_H}
\newcommand{\hr}{ \bar{r}_+}
\newcommand{\new}{\newcommand}
\new{\be}{ \begin{equation}}
\new{\ee}{ \end{equation}}
\new{\bea}{ \begin{eqnarray}}
\new{\eea}{ \end{eqnarray}}
\begin{document}

\begin{flushright}
	   KAIST-CHEP-95/8        \\
\end{flushright}
\begin{center}
 \Large{ The Entropy of a  Quantum Field in a Charged Kerr Black Hole}
\end{center}
 
\vspace{2cm}
   
\begin{center}
  {Min-Ho Lee\footnote{e-mail~:~mhlee@chep6.kaist.ac.kr
  } and Jae Kwan Kim     \\
  Department of Physics, Korea Advanced Institute of 
  Science and Technology \\
 373-1 Kusung-dong, Yusung-ku, Taejon 305-701, Korea.}
\end{center}
\vspace{2cm}
		 
\begin{abstract}
We calculate the entropies  of the system of classical particles and 
a quantum scalar field by using the brick wall method 
in thermal bath in a charged Kerr black hole spacetime.
Their  leading terms 
at Hartle-Hawking temperature $T_H = \kappa/(2 \pi) $ 
are given by $ S_{cl} \approx N \
\ln \left( \frac{A_b}{\epsilon^2} \right)$, 
and $S \approx N' \frac{A_H}{\epsilon^2}$, 
where $A_b$ and $A_H$  are  the area of the box and 
the horizon respectively.
\end{abstract}

\newpage

\section{Introduction}

In 1973 Bekenstein, by comparing the black hole physics with the 
thermodynamics, argued that the black hole entropy is  
proportional to the black hole horizon area \cite{Bekenstein1}. 
Hawking showed that the proportional coefficient  
is $\frac{1}{4}$ by investigating the quantum fields in a 
collapsing black hole space time \cite{Hawking}.
 By using the Euclidean path integral  Gibbons and Hawking  showed that
 the tree-level contribution of the gravitation  action gives 
 the black hole entropy \cite{Gibbon}.
However the exact statistical  origin of the Bekenstein-Hawking entropy 
$S_{bh}$ is unknown.

Recently many efforts have been concentrated to understand the 
 statistical  origin of Bekenstein-Hawking black hole 
entropy \cite{Bekenstein2}.
Frolov and Novikov  argued that the black hole 
entropy can be obtained by identifying 
the dynamical  degrees of freedom of a black hole with 
the states of all fields
which are located inside the black hole \cite{Frolov}. 
Another  approach is to identify the  black hole entrophy $S_{bh}$ 
with the entanglement entropy $S_{ent}$ \cite{Ent}.
Entanglement entropy arises from ignoring the degree of 
freedom of a proper 
region of space.  It is found that the entropy is proportional 
to the area of the boundary (horizon).  However the entanglement 
entropy  is divergent.
Such divergences  also  arise in  the brick wall method  
of t'Hooft \cite{tHooft,Pa1}, who calculated the entropy  
of quantum field
propagating outside the black hole.
The divergences arise from    the density of levels diverges 
due to the infinite shift of frequencies near the horizon.
But It has been shown that the black hole   entropy is proportional to
the area after an appropriate renormalization \cite{Barbon}.
The thermodynamic approach using the heat kernel gives the  same result 
with the brick wall method or the entanglement entropy 
method \cite{optical}.

In this paper  we shall  
investigate the black hole entropy of a scalar field by the brick wall 
method in a charged Kerr   black hole.  We will also study the classical
entropy of particles.
Our result shows that in classical and quantum level the entropies 
are  divergent as the system approaches to the horizon.
The leading entropy of a quantum field in  the Hartle-Hawking state 
is proportional to the area of the event horizon.


\section{The Partition Function of Classical Particle }

Let us consider   a box   containing  N non-interacting  particles
with mass $\mu^2$ described by a Hamiltonian $H (p,x )$ in  
the charged Kerr black hole space-time. We assume that 
the box is rotating with a constant azimuthal angular 
velocity $\Omega_0$.  The line element of the charged Kerr black hole 
spacetime in Boyer-Lindquist coordinates is   
given by \cite{Newmann,Wald} 
\begin{eqnarray}
\nonumber
ds^2 &= & - \left( 
		\frac{ \Delta - a^2 \sin^2 \theta }{\Sigma}
		\right) 
		dt^2 - \frac{2 a \sin^2 \theta ~( r^2 + a^2 - \Delta)}{
		\Sigma } dt d\phi \\
    & &~ + \left[ \frac{(r^2 +a^2 )^2 - \Delta a^2 \sin^2 \theta }{
\Sigma}  \right] \sin^2 \theta d \phi^2 + \frac{\Sigma}{\Delta} dr^2 +
    \Sigma d \theta^2 \\
&\equiv & g_{tt}~ dt^2  + 2 g_{t \phi}~ dt d \phi + 
       g_{\phi \phi }~ d\phi^2 +  g_{rr} ~dr^2 +  
       g_{\theta \theta }~ d\theta^2,
\end{eqnarray} 
where 
\begin{equation}
\Sigma = r^2 + a^2 \cos^2 \theta, ~~~~~  
\Delta = r^2 + a^2 + e^2 - 2 M r,
\end{equation}
and $e,a,$ and $M$ are charge, angular momentum per unit mass, and
mass of the spacetime respectively. 
This spacetime has two Killing vectors: the time like Killing 
vector $\xi^\mu = \left( \partial_t \right)^\mu$ and the axial 
Killing vector $\psi^\mu = \left( \partial_\phi \right)^\mu$.
We assume that $e^2 + a^2 \leq M^2 $.
In this case the charged  Kerr black hole has the event   horizon 
 at $ r_{horizon} = r_+ = M + \sqrt{M^2 - a^2 - e^2 }$  and 
 has the   stationary limit surface    at $ r_0 = M + 
\sqrt{M^2 -  e^2  - a^2 \cos^2 \theta }$. 
Near the event horizon $r = r_+$ the metric behaves like \cite{Thorne}
\begin{equation}
ds^2  \simeq  - ( \kappa \rho)^2 dt^2 + g_{\phi \phi} (r = r_+, \theta)
\left( d \phi - \Omega_H  dt \right)^2 + d\rho^2 + g_{\theta \theta } 
(r = r_+, \theta) d \theta^2,     \label{rindler}
\end{equation}
where $\kappa$ is the surface gravity of the black hole and
$ \rho  = \int \sqrt{g_{rr} } dr$,  
$\Omega_H = \frac{a}{r_+^2 + a^2 }$.
$\Omega_H$ is the angular velocity of the black hole.
This form (\ref{rindler}) is similar to the Rindler metric.

When a system with $N$ non-interacting particles is 
in thermal equilibrium state
at temperature $T = 1/\beta$ and is rotating with a angular 
velocity $\Omega_0$ about $z-$ axis,  the partition function 
$Z_N$  is given by
\begin{equation}
Z_N = (Z)^N,
\end{equation}
where $Z$ is the partition function for one particle:
\begin{equation}
Z = \int  d^3 x d^3 p~ e^{ - \beta ( {\cal E} - \Omega_0 p_\phi )},
\end{equation}
where  ${\cal E}$, which  is the 
energy in $\Omega_0 = 0$ frame and  given by 
$\xi^\mu p_\mu = - p_t = {\cal E}$,
satisfies the following
\begin{equation}
{\cal E} = \frac{1}{- g^{tt}} 
      \left\{ - g^{t \phi} p_\phi \pm 
	 \left[
            (g^{t  \phi} p_\phi )^2 + ( - g^{tt}) ( \mu^2 + 
	    g^{\phi \phi } p_\phi^2 + g^{rr} p_r^2 + g^{\theta \theta } 
	    p_\theta^2 )
         \right]^{1/2} 
     \right\} \equiv  {\cal E}_\pm.      \label{con2}
\end{equation}
Here we used $ p^\mu p_\mu = - \mu^2$. 
${\cal E} - \Omega_0 p_\phi \equiv E$ is the energy of the
particle in rotating frame.
We will take ${ \cal E}_+$ because ${\cal E}_+$ corresponds to 
4-momentum pointing toward future \cite{Misner}.

Note that we must restrict the system to be in the region such that
${g'}_{tt} \equiv  g_{tt} + 2 \Omega_0 g_{t \phi} + 
\Omega_0^2 g_{\phi \phi}
< 0 $. In the region such that $ - {g'}_{tt} >0$ ( called region I) 
the possible points of $p_i$  satisfying $ {\cal E}_+ - 
\Omega_0 p_\phi = E$ 
for a given $E$ are located on the following surface
\begin{equation}
\frac{p_r^2}{g_{rr}} + \frac{p_\theta^2}{g_{\theta \theta}} +
\frac{- {g'}_{tt}}{- \cal D} \left(
      p_\phi + \frac{g_{t \phi }  + 
      \Omega_0 g_{\phi \phi}}{{g'}_{tt}} E 
                             \right)^2 
   = \left( \frac{1}{- {g'}_{tt}} E^2 - \mu^2 \right),   
   \label{ellipsoid}
\end{equation}
which is the ellipsoid,  a compact surface. Here 
$ -{\cal D} = g_{t \phi}^2 - g_{tt} g_{\phi \phi}$. So the 
density of state
$g(E)$ for a given $E$ is finite and the integrations 
over $p_i$  give a  finite value.
But in the region such that $ - {g'}_{tt}<0$ (called region II) 
the possible points of $p_i$  are located on the following surface
\begin{equation}
\frac{p_r^2}{g_{rr}} + \frac{p_\theta^2}{g_{\theta \theta}} -
\frac{ {g'}_{tt}}{- \cal D} \left(    p_\phi + \frac{g_{t \phi }  + 
 \Omega_0 g_{\phi \phi}}{{g'}_{tt}} E         \right)^2 =
- \left( \frac{1}{ {g'}_{tt}} E^2  + \mu^2 \right),
\end{equation}
which is  the hyperboloid,   a non-compact surface. So 
$g(E)$ diverges and the partition function $Z$ diverges. 
Thus we will assume that the box is in the region I.
For example, in the case of $\Omega_0 = 0$ the points 
satisfying ${g'}_{tt} =0$ are on the stationary limit 
surface. The region of the outside (inside) of the stationary 
limit surface corresponds to the region I (II).

The partition function is then given by
\begin{eqnarray}
\nonumber
Z &=& \int_{region~I} d^3 x d^3 p \exp \left\{
        -\beta \left[ ( \Omega  - \Omega_0 ) p_\phi  
	\right.   \right.  \\
     & &   \left.  \left.   + \frac{1}{- g^{tt}}
                    \left[
            (g^{t \phi} p_\phi )^2 + ( - g^{tt}) ( \mu^2 + 
	    g^{\phi \phi } p_\phi^2 + g^{rr} p_r^2 + g^{\theta \theta } 
	    p_\theta^2 )
                    \right]^{1/2} 
              \right]   \right\},
\end{eqnarray}
where $\Omega = - \frac{g_{t \phi}}{g_{\phi \phi}}$.
After the integrations over $p_i$  we get 
\begin{eqnarray}
\nonumber
Z &=& 4 \pi \int_{region ~I}  d^3 x \int_{\mu 
\sqrt{ - {g'}_{tt}}}^\infty d E 
   \frac{ \sqrt{ g_4}}{\sqrt{ { - {g'}_{tt}}}} 
\frac{E}{- {g'}_{tt}} \left( \frac{E^2}{-{g'}_{tt}}  
- \mu^2 \right)^{1/2}
 e^{- \beta E}   \\
 &=&  \frac{4 \pi}{3} \beta \int_{region ~I}  d^3 x
 \int_{\mu \sqrt{ - {g'}_{tt}}}^\infty d E 
   \frac{ \sqrt{ g_4}}{\sqrt{ { - {g'}_{tt}}}} 
 \left( \frac{E^2}{-{g'}_{tt}}  - \mu^2 \right)^{3/2}
 e^{- \beta E},               \label{cpartition}
\end{eqnarray}
where we have integrated by parts.  From this expression 
we easily obtain 
the total number of state $\Gamma_{cl} (E)$ with energy less than $E$ 
\begin{equation}
\Gamma_{cl} (E) = 
   \frac{4 \pi}{3} \int d^3 x 
   \frac{ \sqrt{ g_4}}{\sqrt{ { - {g'}_{tt}}}} 
 \left( \frac{E^2}{-{g'}_{tt}}  - \mu^2 \right)^{3/2},
\end{equation}
which is identical to the result of ref.\cite{Pa2} when $
\Omega_0= a = e = 0$.
This expression also can be  obtained directly by investigating 
Eq.(\ref{ellipsoid}).  

Let us assume that $\Omega_0 \simeq \Omega_H$, 
which is  the angular velocity of the charged Kerr black hole.
In other words the system is co-rotating with the black hole.
We assume that the box is close to the horizon.
Let the radial coordinates of the lower bound and the upper bound of 
the box be $r_+ + h$ for small $h$ and be $L$ respectively.

Then  the leading behavior of partition function $Z$ for $\mu = 0$ 
is  given by 
\begin{eqnarray}
Z    &\approx &    \frac{8 \pi}{ \beta^3}  
\int_{r_+ + h}^L dr d \theta d\phi
\sqrt{ g_{\theta \theta} g_{\phi \phi} } \sqrt{g_{rr}}
\left( \frac{g_{\phi \phi}}{ g_{t \phi}^2 - g_{tt}g_{\phi \phi}}  
\right)^{3/2}   \\
&\approx  & \frac{4 \pi}{ (\kappa \beta )^3 } \frac{A_b}{\epsilon^2},
\end{eqnarray}
where  $A_b$ is the area  of  the box near the horizon .  
$ \epsilon$ is the proper distance from the horizon $r_+$ to $r_+ + h$:
\begin{equation}
\epsilon = \int_{r_+}^{r_+ + h} dr \sqrt{g_{rr}} \approx 
2 \left( \frac{ r_+^2 + a^2 \cos^2 \theta }{ 2 r_+ - 2 M } 
\right)^{1/2} \sqrt{h} 
\end{equation}
for very small $h$.
Therefore as the box approaches to the horizon, the 
partition function $Z$
goes  to be divergent  quadratically in $1/\epsilon^2$.
The particular point is that $\epsilon $ depends on the coordinates
$\theta$.
The partition function of $N$ particles  near the horizon becomes
\begin{equation}
Z_N \approx \left(  \frac{4 \pi}{ (\kappa \beta )^3 } 
\frac{A_b}{\epsilon^2}
\right)^N.
\end{equation}
>From this expression we obtain the leading behavior of  the classical 
entropy 
\begin{equation}
S_{cl}   \approx  N \ln  \left(  \frac{4 \pi}{ (\kappa \beta )^3 } 
\frac{A_b}{\epsilon^2}  \right),  \label{clentropy}
\end{equation}
which diverges logarithmically in $\epsilon$ as the box 
approaches to the 
horizon.

Now let us  reconsider above problem in comoving 
coordinate which co-rotate
with the box. The metric in comoving frame is 
\begin{equation}
ds^2 =  ( g_{tt} + 2 \Omega_0 g_{t \phi} + 
\Omega_0^2 g_{\phi \phi}  ) dt^2
   + 2 ( g_{t \phi } + \Omega_0 g_{\phi \phi} ) dt d \phi'
   + g_{\phi \phi} d {\phi'}^2 +  g_{rr} dr^2 + 
   g_{\theta \theta} d \theta^2,
\end{equation}
where we used $\phi' = \phi - \Omega_0 t$.
In this frame the partition function is given by  
\begin{equation}
Z = \int d^3 x d^3 p e^{- \beta E'}
\end{equation}
where the energy $E'$   satisfies following 
\begin{equation}
E' = \frac{1}{- {g'}^{tt}} 
      \left\{ - {g'}^{t \phi} p_\phi \pm 
	 \left[
            ({g'}^{t \phi} p_\phi )^2 + ( - {g'}^{tt}) ( \mu^2 + 
	    {g'}^{\phi \phi} p_\phi^2 + 
	    {g'}^{rr} p_r^2 + {g'}^{\theta \theta} 
	    p_\theta^2 )
         \right]^{1/2} 
     \right\} \equiv  {E'}_\pm.
\end{equation}
It is easy to show that $ E_+^{'} = {\cal E}_+ - \Omega_0 p_\phi$. Thus 
the partition function give the same value with (\ref{cpartition}).
In this frame the reason that we must restrict a system to be in the 
region I becomes apparent. In the region II the box must 
move greater than 
the velocity of light. So it is unphysical in classical mechanics.
Similar phenomena appears in the quantum field theory case. 
\section{A Rotating  Scalar Field in the Charged Kerr Black Hole }

Let us consider   a minimally coupled scalar field in  
 thermal equilibrium at  temperature $1/\beta$  in the 
charged Kerr black hole spacetime. 
 We assume that the scalar field  is rotating with a constant 
azimuthal angular velocity $\Omega_0$.

For such a equilibrium ensemble of the states  of  the scalar field 
the partition function is given by 
\begin{equation}
Z = \sum_{n_q,m} e^{- \beta ( E_q - \Omega_0  m )n_q },
\end{equation}
where $q$ denotes a quantum state of the field with   energy
$E_q$ and  azimuthal  angular momentum $m$.
The free energy is given by 
\begin{equation}
 \beta F  
 =  \sum_m \int_0^\infty dE  g(E, m) \ln 
       \left( 1 - e^{- \beta( E - m \Omega_0 )}   \right), 
\end{equation}
where $g(E,m)$ is the density of state for a given $E$ and $m$.

To evaluate the free energy we will follow 
the brick wall method  of 't Hooft \cite{tHooft}.
Following the brick wall method we impose a small cut-off $h$
such that 
\begin{equation}
\Phi  (x) = 0 ~~~~{\rm for }~~~ r = r_+ + h.
\end{equation}
To remove the infra-red divergence   we also introduce another 
cut-off  $L \gg r_+$ such that $\Phi (x) = 0$ for $r = L$.

In the WKB approximation with 
$\Phi = e^{- i E t + i m \phi + i S(r, \theta)}$
the Klein-Gordon equation $  ( \Box  - \mu^2 ) \Phi  = 0$ yields 
the constraint \cite{Mann}
\begin{equation}
  p_r^2 = \frac{1}{g^{rr}}  \left[
       - g^{tt} E^2 + 2 g^{t \phi} E m  - g^{ \phi \phi } m^2 
       - g^{ \theta \theta } p_\theta^2 
       - \mu^2      \right],         \label{con1}
\end{equation}
where  $  p_r = \partial_r S$ and $ p_\theta = \partial_\theta S$.
The number of mode with energy less than $E $ and with a fixed $m$ is 
  obtained by integrating  over $p_\theta$ in phase  space
\begin{eqnarray}
  \Gamma (E, m) &=& \frac{1}{\pi} \int d \phi d \theta 
         \int_{r_+ + h}^L dr \int d p_\theta   p_r ( E, m, x)  \\
 \nonumber
	 &=& \frac{1}{\pi} \int d \phi d \theta 
         \int_{r_+ + h}^L dr   \int  d  p_\theta \left[
               \frac{1}{g^{rr}}  \left(
       - g^{tt} E^2 + 2 g^{t \phi} E m  - g^{ \phi \phi } m^2 
       - g^{ \theta \theta } p_\theta^2 
       - \mu^2                   \right)          
						  \right]^{1/2}.
\end{eqnarray}
The   integration over $ p_\theta$ must be carried out over the phase 
space that satisfies $ p_r \ge 0$.


Note that if we identify $m$ as $p_\phi$  then the rearrangement
of the expression (\ref{con1}) yields Eq. (\ref{con2}).  It is 
natural because  WKB approximation  means to treat the 
system as a classical  one.  In the WKB approximation the 
energy $E$  satisfies the constraint (\ref{con1}).
 Thus in the region I,  $E - \Omega_0  m  >  0$. However
 in the region II it is possible that  
 $E - \Omega_0 m  < 0$. (It  is a superradiance mode.) 
 However as in classical case  the geometry of the phase 
 space  for a given $E$ is a non-compact surface.
 Thus the free energy diverges. In fact the surface such that 
 ${g'}_{00} = 0$ is the velocity of light surface (VLS). 
 Outside that surface  a comoving observer must have  a velocity 
 $v_{ob} > 1$ and  move on a spacelike world line. 

In  $\Omega_0  = \Omega_H$ case we can exactly find  the 
position of the  light of velocity surface.  
(In this case the region I corresponds to $ r_+ < r < r_{VLS}$.)
In such a case ${g'}_{tt}$ can be written as
\begin{eqnarray}
{g'}_{tt} &=& g_{tt} + 2 \Omega_H g_{t \phi} + \Omega_H^2 
	         g_{\phi \phi} \\
\nonumber
	  &=& \frac{M^2}{\Sigma}  ( x - \hr)  \left\{
	  \bar{\Omega}_H^2  \sin^2 \theta ~x^3 + \bar{r}_+
	  \bar{\Omega}_H^2 \sin^2 \theta ~x^2    \right.    \\
\nonumber
	 & & ~ + \left.  \left[ -1 + \ome^2 \sin^2 \theta   \left(
	  y^2 + y^2 \cos^2 \theta  + \hr^2    \right)    
	       \right]  x
	  + B     \right\}    \\
	  &\equiv& \frac{M^2}{\Sigma}  ( x - \hr ) 
	   \ome^2 \sin^2 \theta  \left(
	    x^3  + a_1 x^2  + a_2 x + a_3  \right)
\end{eqnarray}
for $\theta \neq 0$, where
$x = r/M, y = a/M, z = e/M, \ome = M \Omega_H, \hr = r_+ /M $, and
\be
\nonumber
B   =  2 \left( 1 - \ome y \sin^2 \theta \right)^2 - \hr + 
 \hr \ome^2 \sin^2 \theta  \left( \hr^2 + y^2  + y^2 \cos^2 \theta 
 \right).
\ee
Then the exact position of the light of velocity surface is given by
\cite{Table}
\be
 r_{light ~of ~velocity} \equiv r_{VLS}   =  2 M \sqrt{
 - Q} \cos \left( \frac{1}{3} \Theta  \right) - \frac{1}{3} a_1 M,
 \label{sol}
 \ee
 where
 \be
 \Theta = \arccos \left(  \frac{R}{\sqrt{- Q^3}} \right)
 \ee
 with
 \be
 Q = \frac{3 a_2 - a_1^2 }{9},~~ 
 R = \frac{9 a_1 a_2 - 27 a_3 - 2 a_1^3 }{54}.
 \ee
 Eq. (\ref{sol})  approximately is given by 
 \be
 r_{VLS} \sim \frac{1}{\Omega_H \sin \theta } - \frac{r_+}{3  },
 \ee
 which is an open, roughly, cylindrical surface.
 For $\theta = 0$ it is always that  ${g'}_{tt} < 0 $ for 
 $r > r_+$.  

 For  extreme charged Kerr  black hole case $M^2 = a^2 + e^2 $ 
 the  position of VLS at $\theta = \frac{\pi}{2} $ is  given by  
\bea
  r &=& M ~~~~~~{\rm for ~}   
  \frac{1}{2} M~\leq a \leq M~ {\rm and }~ a = 0, \\
  r &=& \left( -1 + \frac{M}{a} \right) M  ~~~~~{\rm for ~}  
       0 < a <  \frac{1}{2} M.  
\eea
The second case  corresponds to the extreme  black hole  that
is  slowly rotating and has many charge. (In this case 
 $ e  > \sqrt{3}/2M \approx 0.866 M $). 
In particular in case of $e \leq  
\sqrt{3}/2 M $ ( $a = M$  for $ e = 0$)  
the horizon and  the light of velocity surface  
are at the same position. 
 Therefore  in case of  the extreme black hole with $a \geq 1/2 M $
 it is impossible to consider the brick wall model of 't Hooft. 

Hereafter  we assume   that the outer brick wall is located  
 inside the velocity of light surface, and 
 that the black hole is not extrme.
 About the  location for the outer brick wall 
 (perfectly reflecting mirror)  it was already pointed  
  out in ref.\cite{Thorne} to  remove the singular structure of 
 the Hartle-Hawking vacuum state and to modify it.


 The free energy is then written as 
\begin{eqnarray}
\nonumber
  \beta F  &=& \sum_m \int_{m \Omega_0}^\infty dE  g(E, m) \ln 
       \left( 1 - e^{- \beta( E - m \Omega_0 )}   \right)    \\
\nonumber
       &=& \int_0^\infty dE \sum_m g(E + m \Omega_0 , m) \ln 
       \left( 1 - e^{- \beta E }   \right)             \\
       &=&  - \beta \int_0^\infty dE \frac{1}{e^{\beta E} - 1}
            \int d m \Gamma (E + m \Omega_0, m),
\end{eqnarray}
where we have integrated by parts and we assume that the quantum 
number $m$ is a continuous variable.   
The integrations over $m$ and $p_\theta$  yield 
\begin{equation}
F =   - \frac{4 }{3}  
\int d \phi d \theta \int_{r_+ + h}^L dr
\int_{\mu \sqrt{- {g'}_{tt}}}^\infty dE  
\frac{1}{e^{\beta E} - 1 }
\frac{ \sqrt{g_4}}{\sqrt{ - {g'}_{tt} }}  \left( 
\frac{E^2}{ -  {g'}_{tt} }  - \mu^2  \right)^{3/2 },    
\label{freeenergy}
\end{equation} 
where  
${g'}_{tt}$ can be written as
\begin{equation}
{g'}_{tt} = \frac{{\cal D}}{g_{\phi \phi}}  \left[
	 1 - \left( \Omega - \Omega_0 \right)^2 
	   \frac{g_{\phi\phi }^2}{- {\cal D}}
				    \right].           \label{g00}
\end{equation}
Note that the  form of the free energy is similar to 
Eq. (\ref{cpartition}). 
In particular when $\Omega_0 = a = e = 0$, the 
free energy (\ref{freeenergy}) 
coincides with the expression obtained by 't Hooft \cite{tHooft} and 
it is proportional to the volume of the optical space \cite{optical}.
It is easy to see that the  integrand diverges  as  
$h \rightarrow 0$.

Let $\mu = 0$. For a massless  scalar field the free energy reduces to 
\begin{equation}
\beta F = - \frac{N}{\beta^3 }  \int d \theta d \phi 
\int_{r_+ + h}^L dr \frac{\sqrt{g_4}}{ ( - g^{'}_{tt } )^2  } 
 = - N \int_0^\beta d \tau \int d \theta d \phi 
\int_{r_+ + h}^L dr \sqrt{g_4} 
 \frac{1}{ \beta_{local}^4},            \label{free}
\end{equation}
where $\beta_{local} = \sqrt{ - g^{'}_{tt } } \beta$ is 
the reciprocal of the local Tolman temperature \cite{Tolman}
and $N$ is a constant.
This form  is just the free energy of a gas of massless 
particles at local
temperature $1/\beta_{local}$.
>From this expression (\ref{free})  it is easy to obtain expressions for 
the total energy $U$,  angular momentum  $J$,  and entropy   $S$ of 
a scalar field
\begin{eqnarray}
J &=&   \langle m \rangle = - \frac{1}{\beta} 
\frac{\partial}{\partial \Omega_0}
( \beta F) =     \frac{4 N}{\beta^4}  \int d \theta d \phi 
\int_{r_+ + h}^L dr \frac{\sqrt{g_4}}{ ( - g^{'}_{tt } )^2  }
\frac{ g_{\phi \phi}}{( - {g'}_{tt})}  
\left( \Omega_0 - \Omega \right),  \\
U &=& \langle E \rangle = \Omega_0 J + \frac{\partial}{\partial \beta} 
( \beta F)   =   \frac{N}{\beta^4}  \int d \theta d \phi 
\int_{r_+ + h}^L dr \frac{\sqrt{g_4}}{ ( - g^{'}_{tt } )^2  }
\left[ 
3 + 4 \frac{ \Omega_0 \left( \Omega_0 - \Omega \right)
 g_{\phi \phi } }{( - {g'}_{tt})}  \right],  \\
S &= & \beta^2 \frac{\partial}{\partial \beta } F = 
       \beta ( U - F - \Omega_0 J) = 
       4 \frac{N}{\beta^3 }  \int d \theta d \phi 
        \int_{r_+ + h}^L dr \frac{\sqrt{g_4}}{ ( - g^{'}_{tt } )^2  }.
\end{eqnarray}
Now let us assume that $\Omega_0 \sim \Omega_H$, i.e. the scalar field 
is co-rotating with the black hole. 
Then near the event horizon $r = r_+$,
the leading behavior of the free energy $F$ for small $h$ is 
\bea
\nonumber
\beta F & \approx& -  \frac{N}{\beta^3} \int d \phi d \theta 
\int_{r_+ + h}^L dr
 \sqrt{g_4}  \left( \frac{ g_{\phi \phi} }{ - {\cal D} }  \right)^2  \\ 
\nonumber
 &=& -  \frac{N}{\beta^3} \int d \phi d \theta \int_{r_+ + h}^L dr
\sqrt{g_{\theta \theta} g_{\phi \phi}} \sqrt{g_{rr}}
\left( \frac{g_{\phi \phi}}{g^2_{t \phi} -g_{tt}g_{\phi \phi}}
\right)^{3/2}   \\
&\approx& - N \frac{1}{2 (  \kappa \beta)^3 }  \frac{A_H}{\epsilon^2},
\label{free2}
\eea
where $A_H$ is the area of the event horizon.
The  leading behaviors of the  total  angular momentum $J$, energy 
$U$ and  entropy $S$ are
\begin{eqnarray}
J &=&  0,   \\
 U &=&  3 N \frac{1}{ 2 \kappa^3 \beta^4 }  \frac{A_H}{\epsilon^2},   \\
  S & =& 4 N \frac{1}{2 (  \kappa \beta)^3 }  \frac{A_H}{\epsilon^2}. 
  \label{Entropy}
\end{eqnarray}
The leading behaviors of the thermodynamical 
quantities $U$, and $S$ are 
divergent as $h \rightarrow 0$. But the angular momentum $J$ is $0$.
The divergences arise because   the phase volume 
$\Gamma(E) = \int d m \Gamma(E + m \Omega_0,m) $ diverges as $h$ 
goes to zero.
Actually the phase volume  $\Gamma(E)$ is the same to the classical 
one $\Gamma_{cl}(E)$.

If we take $T$ as the Hartle-Hawking temperature 
$T_H = \frac{ \kappa}{2 \pi}$ 
( In this case the quantum 
state is the Hartle-Hawking vacuum state $|H \rangle$ \cite{Hartle}.) 
the  entropy  becomes
\begin{equation}
S = N' \frac{ A_H}{ \epsilon^2}   \label{result},
\end{equation}
where $N'$ is a new constant.
The entropy of a scalar field diverges quadratically  
in $\epsilon^{-1}$
as the system approaches to the horizon.
Our result (\ref{result}) agrees with the   result calculated 
by 't Hooft \cite{tHooft}.
This fact implies that the leading behavior of entropy  (\ref{result})
is general form.
\section{Discussion}

We have calculated the entropies  of the systems of 
classical particles and a quantum field at thermal 
equilibrium with temperature $T$ in the charged 
Kerr black hole.  
The leading behavior of the entropy of a quantum field  is 
proportional to the area of the event horizon. 
But the classical entropy does not proportional to it.  
Such   leading  forms of the entropies
 can be also  easily calculated by studying the asymptotic 
behavior of the metric (\ref{rindler}) near the horizon. 

Consider a locally co-rotating observer at a  point  near the horizon, 
who carries the following  orthonormal frame
\be
ds^2 = - ( \kappa \rho )^2  dt^2 + g_{\phi \phi}( r = r_+, \theta )
        d {\phi'}^2    + d \rho^2  
     +  g_{\theta \theta} (r = r_+, \theta ) d \theta^2.
\ee
Then the locally measured energy by him is given by 
$E_{loc} = E/(\kappa \rho )  $.
Therefore he will think that the allowed momentum space volume 
is given by 
\be
V_{loc}  = \frac{4 \pi}{3}  \left[ E^2_{loc} - \mu^2 \right]^{3/2}
= \frac{4 \pi}{3} \left[  \frac{E^2}{ ( \kappa \rho )^2 }
- \mu^2 \right]^{3/2},
\ee
and the total phase volume is
\bea
\nonumber
\Gamma (E) &=& \int d \theta d \phi d \rho  
\sqrt{g_{\phi \phi}  g_{\theta \theta}}
~~V_{loc}    \\
\nonumber
&=& \frac{4 \pi}{3} \int  d \theta d \phi  
\sqrt{g_{\phi \phi}   g_{\theta \theta}} \int d \rho  
\left[ \frac{E}{ \kappa \rho } \right]^{3} \\ 
&=& \frac{2 \pi}{3} E^3 \frac{A_H}{ \kappa^3 \epsilon^2} 
\eea 
for $\mu = 0$.
>From this one can obtain the free energy (\ref{free2}) and 
the entropies (\ref{clentropy}), (\ref{Entropy}).
Thus the fundamental reason of divergences is  the infinite number 
of state or the infinite volume of the phase space near the  horizon. 
As in classical case, it can be attributed 
to  the infinite blue  shift of the energy $E$.


In case of  the spherically symmetric black hole,  we need 
the outer brick wall
at $r = L$ to eliminate the infra-red divergence. However 
in the case of the 
rotating black hole we need the outer brick wall at $r = L$, where 
$L < r_{VLS}$,  to prevent a system from having the velocity greater 
than the velocity of light. If not, all thermodynamical quantities are 
divergent.  (It is related to the singularity of the Hartle-Hawking 
state $|H \rangle$ \cite{Thorne}.)
If we consider a rotating system in flat space, such a fact becomes 
more apparent. In WKB approximation the free energy of the 
rotating system 
(see (\ref{freeenergy})) is given by 
\begin{equation}
\beta F = - \frac{N}{\beta^3} \int d \phi dz \int_{region~ I} 
\frac{r}{( 1 - \Omega_0^2 r^2 )^2 } dr
\end{equation}
in the cylindrical coordinates.  As $r \rightarrow 1/\Omega_0 $ ( 
 the velocity of light surface)  the free energy 
diverges.  Outside the  surface of the velocity of light  
(in the region II) the free energy is divergent because 
of the infinite volume of phase space.  Thus we must restrict a 
system in the region I.

The particular point is that in case of the extreme rotating black
such that $ a \geq 1/2 M  $ and $\Omega_0 = \Omega_H$
we can not consider  the brick  
wall model of 't Hooft.  This point is  the different point with 
other extreme black hole like Reissner-Nordstrom  one.



\end{document}